\documentclass[aps, twocolumn, amsfonts, amssymb, pra]{revtex4}

\usepackage{epsfig}

\begin{document}

\title{Dual entanglement measures based on no local cloning and no local deleting}

\author{
Micha{\l} Horodecki\(^{(1)}\), 
Aditi Sen(De)\(^{(2)}\) and Ujjwal Sen\(^{(2)}\)}

\affiliation{\(^{(1)}\)Instytut Fizyki Teoretycznej i Astrofizyki, Uniwersytet 
Gda\'nski, PL-80-952 Gda\'nsk, Poland \\
\(^{(2)}\)Institut f\"ur Theoretische Physik, Universit\"at Hannover, D-30167 Hannover,
Germany}

\begin{abstract}

Impossibility of cloning and deleting of unknown states are important
restrictions on processing of information in the quantum world.
On the other hand, a known quantum state can always be cloned or deleted. 
However if we restrict the class of allowed operations, there will arise 
restrictions on the ability of   cloning and deleting machines. 
We have shown that cloning and deleting 
of known states is in general not possible by local operations. 
This impossibility hints at quantum correlation in the state. 
We  propose dual measures of quantum correlation based on the dual restrictions of 
no local cloning and no local deleting. 
The  measures  are relative entropy distances of the desired states
in a (generally impossible)
perfect local cloning or local deleting process from 
 the best approximate state that is actually obtained by 
imperfect local cloning or deleting  machines. Just like the dual measures of entanglement cost and 
distillable entanglement, the proposed measures are based on
important processes in quantum information.
We discuss their properties.
For the case of pure states, estimations of these two measures are also provided. Interestingly, the entanglement 
of cloning for a maximally entangled state of two two-level systems is not unity.

\end{abstract}

\maketitle

\def\com#1{{\tt [\hskip.5cm #1 \hskip.5cm ]}}

\newcommand{\tr}{{\rm tr}}

\section{Introduction}

In the classical physics description of the world, any two objects are distinguishable. In the quantum world, there 
are some objects which are indistinguishable. In the Hilbert space description 
of the quantum world,  indistinguishable objects are identified with nonorthogonal states, with 
the distinguishable ones being kept as orthogonal.
The existence of nonorthogonal states put restrictions on  the processing of information in quantum world. 

Important such restrictions are the so-called 
no cloning and no deleting theorems \cite{WoottersZurek, Yuen, Pati,Pati2}. 
The no cloning theorem states that nonorthogonal states cannot be copied. 
On the other hand, the no deleting theorem puts restrictions on deleting: Nonorthogonal 
states cannot be deleted \emph{in a closed system}.
We will make it clear, what we mean by a closed system, when we discuss the notion of deleting below
(Section \ref{subsection_no_deleting_global}).  

If one now considers  cloning and  deleting of states shared between separated 
partners, when the partners are allowed to operate only locally, then there are two possible 
ways in which cloning or deleting can  be hindered. First, because of the existence of the no cloning and 
no deleting theorems even for global operations, and second, because of the existence of 
quantum correlations. 
If it possible to wipe out the inefficiency present even in the case of global operations, then 
the remaining inefficiency is solely due to quantum correlations. These could 
then be entanglement measures. However if we have a known state, then by global operations, we can both 
clone and delete. Therefore for a known state, the inefficiency to clone or to delete locally, is solely 
due to quantum correlations. 

In this paper we show that 
the ``distance'' between the possibility of cloning and deleting a known state, shared between separated
partners, and  the impossibility of 
such phenomenon by local actions can reveal the quantum correlation  of the state. 

In this paper, we will primarily deal with bipartite states. However the generalisation to the 
\emph{multipartite case} of all the considerations is straightforward. It is also similarly possible to 
define quantum correlations of \emph{a set of states}, shared between separated partners, or even 
at a \emph{single} location (see \cite{jomey_doi} in this regard). The essential idea is just what has been 
spelled out in the last two paragraphs. We hope to follow that up in a later publication.

Let us stress here that the proposed measures of quantum corelations are defined by basing on 
important phenomena in quantum information.
We hope that this will
be important to understand the resource facet of quantum correlations.

We begin in Section \ref{section_open_closed}, by discussing
no cloning and no deleting for the case of a single system (i.e. 
global operations), mainly to settle down as to what are the operations that are allowed in cloning, and 
what are the ones that are allowed in deleting. 
In Section \ref{sec_entanglement}, we define entangled states, and underline the importance of 
operationally defined measures. Section \ref{sec_cloning_global} deals with no cloning of 
bipartite states under local operations. We show that for distillable states, the entanglement of cloning 
is nonvanishing. The case of bound entangled states \cite{Borda,bound_HHH} is left open.
Section \ref{section_forgetting} concerns the relevant set of operations in case of 
deleting, global and local. 
In Section 
\ref{sec_deleting_global}, we show that 
no deleting of bipartite pure states is not possible by closed local operations.

Since the notions of cloning and deleting are in a sense dual to each other, we call the 
the proposed entanglement measures as dual entanglement measures. The precise definitions of these measures 
is taken up in Section \ref{section_dual}. Some properties of the measures are also proven in the same section. 
In next section (Section \ref{section_bounds}), we obtain bounds on the proposed measures
for the case of pure states of two two-dimensional systems, i.e. of two qubits.  
We obtain that the entanglement of cloning is 
strictly less than unity for the maximally entangled state of 
two qubits (i.e. in \(2 \otimes 2\)).
Note that for almost all 
entanglement measures, this state naturally gives a value of unity. We discuss our results and its possible
extensions in
Section \ref{sec_discu}.

\section{No cloning and no deleting: Open and closed systems}
\label{section_open_closed}

In this section, we briefly discuss the no cloning and no deleting theorems \emph{for a single 
system}, and indicate the importance of open systems and closed systems in such 
considerations. 

We begin with no cloning and underline the fact that it is true in open systems.
In the second subsection, we consider no deleting and the fact that it is true or 
rather relevant only in closed systems.

\subsection{No cloning: Open systems}
\label{subsection_no_cloning_global}

In cloning, the task is to prepare a copy of an unknown state, while 
keeping the input (unknown) copy undisturbed. To be specific, let us suppose that the 
input state is known to be either \(\left|\psi\right\rangle\) or \(\left|\phi\right\rangle\).
We want to obtain two copies of the input, whatever it is. In the process, we allow the environment 
to be possibly changed.

That is, we want to know whether the following transformation is possible: 
\begin{eqnarray}
\label{global_clone}
&& \left|\psi\right\rangle  \left|b \right\rangle \left|e\right\rangle \rightarrow
\left|\psi\right\rangle  \left|\psi\right\rangle \left|e_\psi\right\rangle, \nonumber \\ 
&& \left|\phi\right\rangle  \left|b \right\rangle \left|e\right\rangle \rightarrow
\left|\phi\right\rangle  \left|\phi\right\rangle \left|e_\phi\right\rangle. 
\end{eqnarray}
Here \(\left|b\right\rangle\) is the ``blank'' state, where the second copy is to emerge, 
and \(\left|e\right\rangle\), \(\left|e_\psi\right\rangle\), \(\left|e_\phi\right\rangle\) are 
states of the environment. 
It was shown in Refs. \cite{WoottersZurek,Yuen}, 
that such transformations do not exist in a quantum mechanical 
world, if  \(\left|\psi\right\rangle\) and \(\left|\phi\right\rangle\) are nonorthogonal.
The reason is that quantum mechanical operations are unitary, which must preserve the inner product. 
One can see that the inner product is not preserved in the transformation in Eq. (\ref{global_clone}),
when \(\left|\psi\right\rangle\) and \(\left|\phi\right\rangle\) are nonorthogonal.

Note that in the classical world, any two objects are orthogonal. Consequently  the
transformation in Eq. (\ref{global_clone}) do exist in the classical case
(\(\left|\psi\right\rangle\) and \(\left|\phi\right\rangle\) are orthogonal in that case), a welcome
property allowing us for example to send files over email, while keeping a copy in our computer.

Note also that we allow the environment to be possibly changed after the transformation. So we are considering 
``open systems'' here. The state of the environment, after we have obtained the two copies, may carry 
some information about the state. So, cloning of nonorthogonal states is not possible in open systems. We will come 
back to this point in the next subsection.

\subsection{No deleting: Closed systems}
\label{subsection_no_deleting_global}

In the case of deleting, the task is somewhat dual to that of cloning.
In Refs. \cite{Pati,Pati2}, 
it was shown that no  unitary transformations \cite{not_really} exist which will do the following
transformation in a closed system \cite{Josza, amader1} (cf. \cite{amader2} in this regard), if 
\(\left|\psi\right\rangle\) and \(\left|\phi\right\rangle\) are nonorthogonal:
\begin{eqnarray}
\label{global_delete}
&& \left|\psi\right\rangle  \left|\psi \right\rangle  \rightarrow
\left|\psi\right\rangle  \left|0\right\rangle, \nonumber \\ 
&& \left|\phi\right\rangle  \left|\phi \right\rangle  \rightarrow
\left|\phi\right\rangle  \left|0\right\rangle.
\end{eqnarray}
Here \(\left|0\right\rangle\) is any fixed state. 
This is the no deleting theorem.

Again, it is possible to delete in the classical world, where we only have orthogonal 
objects. This may for example be useful in a 
reversible (i.e. without polluting the environment) defragmentation of the discs in our 
(classical) computers.

Note here that in the case of deleting, we do not allow leakage of information 
from the system into the environment. 
Otherwise, one can trivially delete a copy of the state by throwing it out of the system. 
The point is that even then, the information in the deleted copy must be somewhere  in the world. 
Speaking in the language of the 
Hilbert space formalism of quantum mechanics, we do not allow ``tracing out'' as a 
valid operation. This is what we mean, when we say that we consider ``closed systems''. On the 
other hand, ``open systems'' are those in which we \emph{do} allow tracing out as a valid operation
(along with other quantum mechanically valid operations).

The no deleting theorem is thus relevant only in closed systems. In contrast, the no cloning theorem 
is true even in open systems.

\section{Entanglement}
\label{sec_entanglement}

The restrictions of no cloning and no deleting
in quantum mechanics comes from the superposition principle in quantum mechanics. 
This is because, it is the 
superposition principle that leads to the existence of nonorthogonal  states.
The same superposition principle  gives rise to the existence of entanglement.

In the classical physics description of the world, if a system (in a pure state) can 
be divided into two subsystems, then 
the sum of the information of the subsystems makes up the complete information in the whole
system. This is no longer true in the quantum formalism as there exist entangled states, for example the 
singlet \(\frac{1}{\sqrt{2}}(\left|01\right\rangle - \left|10\right\rangle)\), where 
\(\left|0\right\rangle\) and \(\left|1\right\rangle\) are two orthogonal states.

In general, given a bipartite system, shared between two partners called Alice (A) and Bob (B), the most 
general state that they can create while acting locally and communicating over a classical channel (the set of 
operations being called LOCC (local operations and classical communication)) is 
\begin{equation}
\label{sep}
\sum_i p_i \varrho_A^i \otimes \varrho_B^i,
\end{equation}
where \(\varrho_A^i\) is defined over \({\cal H}_A\), the Hilbert space of the system which is with Alice, and 
similarly \(\varrho_B^i\) is defined over \({\cal H}_B\), the Hilbert space of the system with
 Bob. The \(p_i\)'s are a set of probabilities.
Preparation of more general states require using a quantum channel. Such states possess quantum correlations, and 
can be used for nonclassical applications, for example in teleportation \cite{tele}.
States of the form as in Eq. (\ref{sep}) are called separable states, while the ones that are not of that form are 
called entangled. Entangled states do exist, the singlet state being an example.

Since entanglement is being viewed as a resource, and since it is 
not as yet well-understood, it is important to quantify it in as many ways as possible. 
There have been several very fruitful attempts to quantify entanglement
\cite{IBMhuge, VPRK, VP,
 Vidal-mon2000, Horodecki_limits, qic}. Two important examples of entanglement 
measures are the entanglement cost  \cite{IBMhuge,HHT} and distillable entanglement \cite{IBMhuge,
Rains}. Both these measures are defined operationally and are
 in a sense dual to each other. The entanglement cost (\(E_C\)) of a bipartite state \(\varrho_{AB}\) is 
the asymptotic rate of singlets required to prepare, under LOCC operations, the state \(\varrho_{AB}\) with 
arbitrarily good fidelity. On the other hand, the distillable entanglement (\(E_D\)) of \(\varrho_{AB}\)
is the asymptotic rate of singlets of arbitrarily good fidelity that can be obtained from \(\varrho_{AB}\), by 
LOCC operations. For more objective definitions, see Refs. \cite{HHT} and \cite{Rains} respectively.

Such operationally meaningful entanglement measures are important to understand the resource 
perspective of entanglement. For example, to teleport a qubit (a two-dimensional quantum system) exactly,
one requires a singlet. Given a state \(\varrho_{AB}\) which is not a singlet, we will like to know 
how good it is for teleportation. If we are interested to find out how good an arbitrary qubit can be 
teleported, the relevant quantity is the so-called teleportation fidelity \cite{no_bound_ent-tele}.
However, if we are earnest that the qubit must be teleported (almost) exactly, then the 
relevant quantity is distillable entanglement (or some finite-copy variation of it \cite{Hardy,JP}).

Below (in Section \ref{section_dual}) 
we define two operationally motivated entanglement measures, which are again in a sense 
dual to each other.

\section{No local cloning for entangled states}
\label{sec_cloning_global}

In the case when all (quantum mechanically valid) operations are allowed (this is the case considered 
in Section \ref{subsection_no_cloning_global}), 
a single state can always be cloned. One simply takes the blank copy to be the same as the \emph{known} 
state to be copied (see Eq. (\ref{global_clone})).

Consider however the case where two separated parties, Alice and Bob, are given the task of 
cloning 
the  \emph{known} bipartite state \(\varrho_{AB}\), and the 
allowed operations are restricted to LOCC. Since Alice and Bob are separated, they can at most 
prepare separable blank states for the cloning process. 
Is it then possible 
to produce two copies of \(\varrho_{AB}\)?

Let us make the considerations more formal. Let \(\varrho_{AB}\) be a bipartite state defined on the 
Hilbert space \({\cal H}_A \otimes {\cal H}_B\). This state is given 
to Alice and Bob, who are in two far-apart laboratories. Alice and Bob locally prepare  a state 
\(\varrho^{b}_{A^{'}B^{'}}\) (defined on the Hilbert space 
\({\cal H}_A^{'} \otimes {\cal H}_B^{'}\)), which is to act as the ``blank'' state. Of course, we 
have to take \(\dim {\cal H}_A = \dim {\cal H}_{A^{'}}\), and 
\(\dim {\cal H}_B = \dim {\cal H}_{B^{'}}\). The \(A^{'}\) part of \(\varrho^b_{A^{'}B^{'}}\) is with 
Alice, and the \(B^{'}\) part is with Bob.
This blank state must be 
a separable state, as it is prepared locally (i.e. by LOCC) by Alice and Bob. Note that 
the state \(\varrho_{AB}\) is known to Alice and Bob, and so the blank state can depend on \(\varrho_{AB}\). 
Now the task for Alice and Bob, is to act locally (i.e. Alice acting quantum mechanically on 
\(AA^{'}\), and Bob acting quantum mechanically on \(BB^{'}\), and communicating over a classical channel) 
on the state 
\begin{equation}
\label{debu}
\varrho_{AB} \otimes \varrho^b_{A^{'}B^{'}}
\end{equation}
and produce a state \(\eta_{ABA^{'}B^{'}}\), such that 
\begin{equation}
\label{equal_copy}
\tr_{AB} \eta_{ABA^{'}B^{'}} = \tr_{A^{'}B^{'}} \eta_{ABA^{'}B^{'}} = \varrho_{AB}.
\end{equation}
This is what we mean when we say that the task of Alice and Bob is to prepare two copies of \(\varrho_{AB}\).
In particular, \(\eta_{ABA^{'}B^{'}}\) is \emph{not} needed to be \(\varrho_{AB} \otimes \varrho_{A^{'}B^{'}}\).
Also we will often not explicitly state that the allowed operations are LOCC. See Fig. 1.

\begin{figure}[ht]
\begin{center}
\unitlength=0.5mm
\begin{picture}(250,100)(0,0)
\thicklines
\put(40,5){B}
\put(57,5){\(B^{'}\)}

\thinlines
\put(35,-5){\framebox(30,30)}
 

\put(35,55){\framebox(30,30)}

\thicklines
\put(42,65){\line(0,-1){50}}
\put(59,65){\line(0,-1){50}}

\put(40,69){A}
\put(57,69){\(A^{'}\)}

\thicklines
\put(42,65){\circle*{2}}
\put(59,65){\circle*{2}}

\thicklines
\put(42,15){\circle*{2}}
\put(59,15){\circle*{2}}

\put(25, 36){\(\rho_{AB}\)}

\put(65, 36) {\(\rho_{A^{'}B^{'}}^{b}\)}


\put(100,5){B}
\put(117,5){\(B^{'}\)}

\thinlines
\put(95,-5){\framebox(30,30)}
 

\put(95,55){\framebox(30,30)}

\thicklines
\put(102,65){\line(0,-1){50}}
\put(119,65){\line(0,-1){50}}

\put(100,69){A}
\put(117,69){\(A^{'}\)}

\thicklines
\put(102,65){\circle*{2}}
\put(119,65){\circle*{2}}

\put(102,65){\line(1,0){17}}
\put(102,15){\line(1,0){17}}

\thicklines
\put(102,15){\circle*{2}}
\put(119,15){\circle*{2}}

\put(125, 36){\(\eta_{AA^{'}BB^{'}}\)}

\thicklines
\put(75,46){\vector(1,0){17}}

\end{picture}
\end{center}
\caption{The left hand side of the diagram shows the initial states shared by Alice and Bob. 
\(\rho_{AB}\) is the state shared by them initially, and this is the state that we 
want to clone locally, and \(\rho_{A^{'}B^{'}}^{b}\) is the separable blank state, shared
by Alice and Bob. After the cloning operation has been performed locally by Alice and Bob, they will get 
some four party state \(\eta_{AA^{'}BB^{'}}\), depicted in the right hand side of the diagram.
 If local cloning is possible, then after some local operations, we will obtain a state 
\(\eta_{AA^{'}BB^{'}}\), such that 
 \(\tr_{AB} \eta_{AA^{'}BB^{'}}= \tr_{A^{'}B^{'}} \eta_{AA^{'}BB^{'}} =  \rho_{AB}\). We have shown that local cloning 
of  distillable states is not possible.}
\label{bou-er_kirti} 
\end{figure}
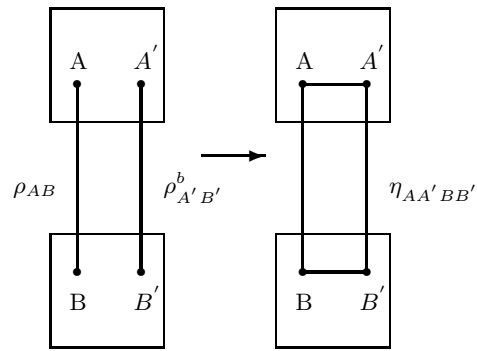

In the case when \(\varrho_{AB}\) is a separable state, 
\(\eta_{ABA^{'}B^{'}}\) can indeed be just \(\varrho_{AB} \otimes \varrho_{A^{'}B^{'}}\),
as the blank state can then be prepared as the state \(\varrho_{A^{'}B^{'}}\).
So separable states can be cloned locally.

What if the state \(\varrho_{AB}\) is entangled? It may seem that the statement ``local cloning of entangled states 
is not possible'' follows 
directly from the famous mantra in quantum communication, ``entanglement cannot 
increase under LOCC''. Maybe it is true, but 
here we are only able to prove that 
local cloning of ``distillable states'' is not possible. Distillable states are 
ones for which the distillable entanglement (\(E_D\)) (see Section \ref{sec_entanglement})
is strictly greater than 0. There exist states, the bound entangled states, which are 
entangled and yet have \(E_D = 0\) \cite{Borda,bound_HHH}.

If \(\varrho_{AB}\) is distillable, then one cannot 
clone it locally (i.e. produce two copies locally).
This follows from the fact that distillable entanglement 
does not increase under LOCC (by the definition of \(E_D\)), and that for a state \(\eta_{ABA^{'}B^{'}}\) (in
the \(AA^{'}:BB^{'}\) cut), 
one has (again by the very definition of \(E_D\))
\begin{equation}
\label{distill_prop}
E_D (\eta_{ABA^{'}B^{'}}) \geq E_D (\tr_{AB}\eta_{ABA^{'}B^{'}}) + E_D (\tr_{A^{'}B^{'}}\eta_{ABA^{'}B^{'}}).
\end{equation}
Consider 
the distillable entanglement, in the 
\(AA^{'}:BB^{'}\) cut, 
of the input \(\varrho_{AB} \otimes \varrho^b_{A^{'}B^{'}}\) and compare it with that of the output 
\(\eta_{ABA^{'}B^{'}}\). 
We have 
\begin{equation}
E_D (\varrho_{AB} \otimes \varrho^b_{A^{'}B^{'}}) = E_D (\varrho_{AB}),
\end{equation}
since \(\varrho^b_{A^{'}B^{'}}\) is a separable state. On the other hand, 
if \(\varrho_{AB}\) can be cloned locally, we have (using relation (\ref{distill_prop}))
\[
E_D (\eta_{ABA^{'}B^{'}}) \geq 2E_D (\varrho_{AB}).
\]
Therefore, whenever \(\varrho_{AB}\) has nonzero distillable entanglement, 
one obtains (in the \(AA^{'}:BB^{'}\) cut)
\[
E_D(\varrho_{AB} \otimes \varrho^b_{A^{'}B^{'}}) < E_D (\eta_{ABA^{'}B^{'}}).
\]
This is a contradiction as \(\eta_{ABA^{'}B^{'}}\) was obtained from 
\(\varrho_{AB} \otimes \varrho^b_{A^{'}B^{'}}\) by LOCC, and \(E_D\) does not 
increase under LOCC.
The contradiction proves that one cannot locally clone distillable states.

\subsection{The case of bound entangled states}

Although the case of bound entangled states is left open, 
we show that under some assumptions, the case can be resolved.

Let us consider an entanglement measure \(E\) of bipartite states, that  satisfies the following inequality:
\begin{equation}
\label{conjecture_eq}
E_{AA^{'}: C}(\xi) \geq E(\mbox{tr}_{A^{'}}\xi) + E(\mbox{tr}_{A}\xi) .
\end{equation}
Here \(\xi_{AA^{'}C}\) is a three-party state and for example, \(E_{AA^{'}: C}(\xi)\) denotes the entanglement 
(as quantified by \(E\)) of \(\xi\) in the \(AA^{'}:C\) cut. 
An example of an entanglement measure which satisfies the inequality (\ref{conjecture_eq}), is  ``squashed
entanglement'',  introduced very recently \cite{Winter} (see also \cite{Alicki}).

Going back to our problem of local cloning, and the four party state \(\eta_{AA^{'}BB^{'}}\),
 we can write, using Eq. (\ref{conjecture_eq}),  
\begin{equation}
\label{35}
E_{AA^{'}: BB^{'}}(\eta) \geq E_{A:BB^{'}}(\mbox{tr}_{A^{'}}\eta) + E_{A^{'}:BB^{'}}(\mbox{tr}_{A}\eta).
\end{equation}

Now, as \(E\) is a measure of entanglement,
\begin{equation}
\label{37}
E_{AA^{'}: BB^{'}}(\varrho_{AB} \otimes \varrho_{A^{'}B^{'}}^{b}) = E(\varrho_{AB}),
\end{equation}
and 
\begin{eqnarray}
E_{A:BB^{'}}(\mbox{tr}_{A^{'}}\eta) &\geq & E(\mbox{tr}_{A^{'}B^{'}}\eta) \nonumber \\
  E_{A^{'}:BB^{'}}(\mbox{tr}_{A}\eta) &\geq & E(\mbox{tr}_{AB}\eta). \nonumber
\end{eqnarray}
If we assume that \(\varrho_{AB}\) can be cloned locally, we have that 
\[
E_{A:BB^{'}}(\mbox{tr}_{A^{'}}\eta) +   E_{A^{'}:BB^{'}}(\mbox{tr}_{A}\eta)
\geq 2 E(\varrho_{AB}),
\]
from which, using the inequality (\ref{35}), we have 
\begin{equation}
\label{36}
E_{AA^{'}: BB^{'}}(\eta) \geq 2 E(\varrho_{AB}).
\end{equation}


Now \emph{if} \(E(\varrho_{AB})\) is strictly positive, then we have (using relations (\ref{37}) and (\ref{36}))
\[
E_{AA^{'}: BB^{'}}(\varrho_{AB} \otimes \varrho_{A^{'}B^{'}}^{b}) 
> E_{AA^{'}: BB^{'}}(\eta).
\]
This is a contradiction, as entanglement cannot increase under LOCC. Then it follows that 
local cloning of the state \(\varrho_{AB}\) is not possible, assuming that there exists an entanglement 
measure \(E\) that satisfies relation (\ref{conjecture_eq}) and that \(E(\varrho_{AB}) > 0\).

Squashed entanglement \cite{Winter} satisfies the inequality (\ref{conjecture_eq}).
However although squashed entanglement is positive for distillable states, 
it is not yet known whether it is nonzero for bound entangled states.
Hence the possibility of local cloning for bound entangled states remain unresolved,  
even with the consideration of squashed entanglement.

Our task is done if we can find an entanglement measure 
that satisfies the relation (\ref{distill_prop}) \emph{or} 
the relation (\ref{conjecture_eq}), \emph{and} also is nonzero for the bound entangled states. 
Relative entropy of entanglement \cite{VPRK,VP}, defined as 
\begin{equation}
\label{relent_def}
E_R(\varrho_{AB}) = \min\limits_{\sigma_{AB} \in {\cal S}} S(\varrho_{AB}|\sigma_{AB}),
\end{equation}
where \(S(\varrho|\sigma) = \tr(\varrho \log_2 \varrho - \varrho \log_2 \sigma)\) and 
\({\cal S}\) is the set of all separable states on \({\cal H}_A \otimes {\cal H}_B\),
is known to be nonzero for bound entangled states. This is because 
the set of all separable states is a closed set, and so any state outside it (the bound entangled states
being among them) cannot be on the boundary of the separable states. 
However we do not know whether \(E_R\) satisfies 
relation (\ref{distill_prop}) or (\ref{conjecture_eq}).


Note that it is very interesting to find whether bound entangled states can be cloned locally. 
It would probably 
reveal at least a part of the not-well-understood physical interpretation of bound entangled states.

\section{Measuring and then forgetting the outcome
is not a closed operation}
\label{section_forgetting}

As we have already mentioned (in Section \ref{subsection_no_deleting_global}), 
in considerations of deleting, the relevant set of operations are the set of closed operations. Unitary 
operations are of course closed operations. Such operations do not leak information into the environment.
The other quantum mechanically valid operation is measurement. Measurement however is not closed. The 
outcome of the measurement is leakage into the environment. But it may seem that performing a measurement, and then 
forgetting the outcome is a closed operation. In this section, we show that this is not the case. Measuring and 
forgetting the outcome leads to leaking of information about the system into the environment, and hence is an open 
operation. Thus the only operation that is closed is the unitary operation. 

When there are several   partners in separated laboratories, closed operations mean performing local unitaries. 
Classical communication is an open operation. Moreover, since there are no measurements performed (by the
partners in their local laboratories), there is no need for a classical communication, as the only 
information to be communicated (during any local 
protocol) is the outcome of any measurement performed.

We now show that measuring and forgetting the outcome is not a closed operation. 
Suppose that a measurement is performed in the 
basis \(\{\left|0\right\rangle, \left|1\right\rangle\}\),
on the state \(\left|\alpha \right\rangle = a\left|0\right\rangle
+ b \left|1\right\rangle\) (\(\left|0\right\rangle\) and \(\left|1\right\rangle\) are 
orthogonal, and \(|a|^2 + |b|^2 =1\)). 
Forgetting the outcome leads to the state 
\(|a|^2 \left|0\right\rangle \left\langle 0 \right| + |b|^2\left|1\right\rangle \left\langle 1 \right|\).
That is, the transformation is 
\begin{equation}
\label{oof}
\left|\alpha \right\rangle \left\langle \alpha \right| \rightarrow
\left|0\right\rangle \left\langle 0 \right|\left|\alpha\right\rangle \left\langle \alpha \right|
\left|0\right\rangle \left\langle 0 \right| +
\left|1\right\rangle \left\langle 1 \right|\left|\alpha\right\rangle \left\langle \alpha \right|
\left|1\right\rangle \left\langle 1 \right|. 
\end{equation}

Any transformation, which is of the form 
\begin{equation}
\label{ek}
\left|\alpha \right\rangle \left\langle \alpha \right| \rightarrow
\sum_i A_i\left|\alpha \right\rangle \left\langle \alpha \right| A^\dagger_i, 
\end{equation}
where \(\sum_i A_i^\dagger A_i\) is the identity operator on the Hilbert space of 
\(\left|\alpha \right\rangle\), 
is equivalent to the unitary operator, on the system \(s\) (i.e. the part \(\left|\alpha \right\rangle\))
along with an environment \(e\), that effects the following transformation:
\begin{equation}
\label{dui}
\left|\alpha \right\rangle_s \left|0 \right\rangle_e 
\rightarrow \sum_i (A_i \left|\alpha \right\rangle)_s \left|i \right\rangle_e.
\end{equation}
Tracing out the environment from Eq. (\ref{dui}), one regains Eq. (\ref{ek}).
Here the \(\left|i\right\rangle\)'s are orthonormal states of the environment, and 
the \(A_i\)'s are operators defined on the Hilbert space of 
\(\left|\alpha \right\rangle\).

So the transformation in Eq. (\ref{oof}) is equivalent to the unitary operator that transforms
\[
\left|\alpha \right\rangle_s \left|0 \right\rangle_e 
\rightarrow  \left|0 \right\rangle_s \left\langle 0 | \alpha \right\rangle \left|0 \right\rangle_e
+ \left|1 \right\rangle_s \left\langle 1 | \alpha \right\rangle \left|1 \right\rangle_e.
\]
However the right hand side is just 
\[
a \left|0 \right\rangle_s \left|0 \right\rangle_e + b \left|1 \right\rangle_s \left|1 \right\rangle_e,
\]
so that, after the measurement is completed and after the observer forgets the outcome,
 the local state of the environment is  
\(|a|^2 (\left|0\right\rangle \left\langle 0 \right|)_e + |b|^2 (\left|1\right\rangle \left\langle 1 \right|)_e\), 
which 
is just the same as the local state of the system, and which \emph{does}
contain information about the initial state \(\left|\alpha\right\rangle\) of the system. 
So as the observer (who performs the measurement) forgets
the outcome
 of the measurement, the environment also forgets the \emph{outcome}. 
However information about the initial \emph{state} of the system (i.e. about 
\(\left|\alpha \right\rangle = a\left|0\right\rangle
+ b \left|1\right\rangle\)) is still in the environment. This is what we wanted to show: 
The operation of a measurement and subsequent forgetting of the outcome is not closed.

\section{No local deleting for entangled states}
\label{sec_deleting_global}


In this section, we will consider deleting of entangled states. 




Consider the case 
when separated partners Alice and Bob are given the task of locally deleting a 
copy from two given copies of a state \(\varrho_{AB}\). So Alice and Bob are 
given the state 
\begin{equation}
\label{arai}
\varrho_{AB} \otimes \varrho_{A^{'}B^{'}},
\end{equation}
where \(AA^{'}\) is with Alice and \(BB^{'}\) is with Bob.
Their task is to  locally (i.e. Alice acting on \(AA^{'}\) and Bob acting on \(BB^{'}\))
transform the given state into \(\zeta_{ABA^{'}B^{'}}\) such that 
\begin{equation}
\label{poune-tin}
\tr_{AB}\zeta_{ABA^{'}B^{'}} 
\end{equation}
is a separable state and 
\begin{equation}
\label{tin}
\tr_{A^{'}B^{'}}\zeta_{ABA^{'}B^{'}} = \varrho_{AB}.
\end{equation}
This may seem to be the easiest thing in the world. One just gets rid of the \(A^{'}B^{'}\) part of the 
input (i.e. one just throws out \(\varrho_{A^{'}B^{'}}\)) and produces any separable state by LOCC.

However as we have already stressed, the notion of deleting is relevant only for closed operations. 
And from section \ref{section_forgetting}, it follows that the allowed operations in that situation 
are only local unitaries. Let us first show that deleting of entangled states is at least possible under 
global unitaries (Section \ref{subsec_delete_global}). 
We will then show that for any bipartite entangled pure state, deleting by 
closed local operations (i.e. by local unitaries) is not possible (Section \ref{subsec_delete_local}).

Before moving on to these proofs, note that the requirement 
that \(\mbox{tr}_{AB} \zeta_{ABA^{'}B^{'}}\) is a separable state,
is a vital element in 
the considerations of local deleting.  Intuitively, in a local deleting process, we are trying to 
delete the shared quantum correlations between the two parties in the second copy, by local 
closed operations. Thus it is natural that we want the second copy, after deletion  has been performed on it,
to be separable.

\subsection{Deleting of entangled states by closed global operations is possible}
\label{subsec_delete_global}

In this subsection, we show that by global unitaries, it is possible to delete entangled states.

Let us first consider the case of pure states. For the case of 
\emph{known pure} bipartite states, deleting a copy from two given copies of any state, 
is possible when closed global operations are allowed. 
Suppose Alice and Bob are given two copies of \(\left|\psi\right\rangle_{AB}\). 
Then there obviously exists a (possibly global) unitary that implements 
\[
\left|\psi\right\rangle_{AB} \left|\psi\right\rangle_{A^{'}B^{'}} \rightarrow 
\left|\psi\right\rangle_{AB} \left|0\right\rangle_{A^{'}}\left|0\right\rangle_{B^{'}},
\]
and the deletion is done.

Moving now to the general case, consider the situation where Alice and Bob are given two copies of \(\varrho_{AB}\),
i.e. they are given the state 
\[
\varrho_{AB} \otimes \varrho_{A^{'}B^{'}}.
\]
They want to delete a copy, while keeping the second copy, in the sense stated above (see Eqs.
(\ref{arai}), (\ref{poune-tin}), (\ref{tin})), by using 
closed global operations. Let us write the spectral decomposition of \(\varrho_{A^{'}B^{'}}\):
\[
\varrho_{A^{'}B^{'}} = \sum_i p_i (\left|\psi_i\right\rangle \left\langle \psi_i \right|)_{A^{'}B^{'}}.
\]
Here \(p_i\)'s are the probabilities, and \(\left|\psi_i\right\rangle\)'s are orthonormal. Since 
\(\varrho_{A^{'}B^{'}}\) is an arbitrary state, the \(\left|\psi_i\right\rangle\)'s may be entangled.

To perform the deleting, Alice and Bob does nothing to the \(AB\) part, while applies the (possibly global)
unitary operator  to the \(A^{'}B^{'}\) part that implements the following transformations for all \(i\):
\[
\left|\psi_i\right\rangle_{A^{'}B^{'}} \rightarrow \left|i_k\right\rangle_{A^{'}} \otimes
\left|i_l\right\rangle_{B^{'}}.
\] 
Here the set \(\{\left|i_k\right\rangle_{A^{'}} \otimes
\left|i_l\right\rangle_{B^{'}}\}_{kl}\)
forms an orthonormal product basis (not necessarily biorthogonal).

After the operation, the output is 
\[
\varrho_{AB} \otimes \sum_i p_i (\left|i_k\right\rangle\left\langle i_k\right|)_{A^{'}} \otimes
(\left|i_l\right\rangle\left\langle i_l\right|)_{B^{'}}.
\]
The \(AB\) part is left intact, while the \(A^{'}B^{'}\) part is now separable. And so we have been able 
to delete one copy of two copies of an arbitrary bipartite state, using closed global operations.

\subsection{Deleting of pure entangled states by closed local operations is not possible}
\label{subsec_delete_local}

We will now show that deleting of pure entangled states, in the sense described above (see Eqs.
(\ref{arai}), (\ref{poune-tin}), (\ref{tin})), is not possible under closed local operations.

We give the proof for two qubit pure states. The case of higher dimensions is similar. Suppose therefore 
that two copies of the two-qubit state \(\left|\psi\right\rangle_{AB} = a\left|00\right\rangle_{AB}
 + b \left|11\right\rangle_{AB} \) are given to Alice and Bob. (\(\left|\psi\right\rangle_{AB}\) 
is written in Schmidt decomposition, so that \(a\) and \(b\) are positive numbers (with \(a^2 + 
b^2 =1\)) and \(\left|0\right\rangle\) and \(\left|1\right\rangle\) are orthogonal.)
So the input is 
\begin{equation}
\label{char}
\left|\psi\right\rangle_{AB} \otimes \left|\psi\right\rangle_{A^{'}B^{'}}.
\end{equation}
Suppose now that  deletion is possible. 
The input in Eq. (\ref{char}) is a pure state, and so after 
closed local operations (i.e. unitaries over \(AA^{'}\) and \(BB^{'}\)),
 the final state must remain pure. 
Moreover, the \(AB\) part of the final state must be in the pure state \(\left|\psi\right\rangle\), as we assume
that deletion is possible. So the final state must be a tensor product of the 
pure state \(\left|\psi\right\rangle\) in the \(AB\) part, and another pure state in the \(A^{'}B^{'}\) part. 
Since we assume that deletion is possible, this state of the \(A^{'}B^{'}\) part in the final state must be a 
pure product state.
Therefore after closed local operations, 
the  total state must be transformed to 
\begin{equation}
\label{panc}
(a\left|00\right\rangle_{AB}
 + b \left|11\right\rangle_{AB}) \otimes \left|0^{'}\right\rangle_{A^{'}} \otimes \left|0^{''}\right\rangle_{B^{'}},  
\end{equation}
where 
\(\left|0^{'}\right\rangle\) and \(\left|0^{''}\right\rangle\) are any two states of \({\cal H}_{A^{'}}\)
and \({\cal H}_{B^{'}}\) respectively. 
But the number of nonzero Schmidt coefficients in the state in Eq. (\ref{char}) is four (if 
\(\left|\psi\right\rangle\) is an entangled state), while there are just two 
nonzero Schmidt coefficients in the state in Eq. (\ref{panc}). This cannot happen under a local unitary 
transformation. (Similar logic is obviously true for higher dimensions.)
The contradiction proves that local 
deletion of pure bipartite entangled states is not possible.

The question as to whether mixed bipartite entangled states can be deleted by closed local operations 
is left open. However if we require that in local deleting of bipartite states, the output must be of the 
form \(\varrho_{AB} \otimes \varrho^{sep}_{A^{'}B^{'}}\), where \(\varrho^{sep}_{A^{'}B^{'}}\) is 
some separable state (see Eqs. (\ref{arai}), (\ref{poune-tin}), 
and (\ref{tin})), then we can show that for distillable states, local 
deleting is not possible. Suppose therefore that the two input copies 
\(\varrho_{AB} \otimes \varrho_{A^{'}B^{'}}\), is by closed local operations 
(in the \(AA^{'}:BB^{'}\) cut) taken to a state of the form 
\(\varrho_{AB} \otimes \varrho^{sep}_{A^{'}B^{'}}\). Then we have (in the \(AA^{'}:BB^{'}\) cut)
\[
 E_D(\varrho_{AB} \otimes \varrho_{A^{'}B^{'}})
=  E_D(\varrho_{AB} \otimes \varrho^{sep}_{A^{'}B^{'}}),
\]
as distillable entanglement is invariant under local unitary operations.
The left hand side is \(\geq 2E_D(\varrho_{AB}) \) while the right hand side equals 
\(E_D(\varrho_{AB}) \). This is a contradiction whenever \(E_D(\varrho_{AB})  > 0\). For 
bound entangled states \cite{Borda, bound_HHH}, the problem is again open. Let us stress that 
the problem of whether local deleting of entangled states is possible is open for all mixed
states. We have answered the question (in the negative) for distillable states, only for 
local deleting in a restricted sense.

\section{The dual measures}
\label{section_dual}

In the previous sections, we have seen that cloning and deleting of a \emph{known} bipartite entangled 
state is in general  not possible under local operations. Let us again stress that the meaning of these 
statements is not very obvious, and that they are elucidiated in Sections 
\ref{sec_cloning_global} and \ref{sec_deleting_global}
 respectively. In 
particular, see Eqs. (\ref{debu}), (\ref{equal_copy})
for the meaning of local cloning and Eqs. (\ref{arai}), (\ref{poune-tin}), (\ref{tin})
for that of local deleting.

We have also seen that it is possible to clone and delete a \emph{known} state under global operations (see
at the begining of Section \ref{sec_cloning_global} and Section \ref{subsec_delete_global}).

It is interesting to find the reason for such restrictions in the case of local operations. 
It seems that the possibility of cloning and deleting in the case of global operations, and 
the impossibility of such events under local operations indicates that the ``amount'' of such impossibility for 
a certain state
in the case of local operations will reveal the amount of quantum correlations in that state.

In this section, we show just that. We obtain two measures of entanglement from the ``amount'' of 
the impossibility of local cloning and local deleting respectively. Since the process of 
local cloning is in a sense dual to that of local deleting, we say that these entanglement measures are 
dual to each other. This is just as for distillable entanglement and entanglement cost (see Section 
\ref{sec_entanglement}).

\subsection{The entanglement of cloning}
\label{subsec_entclo}

\subsubsection{The definition}


Let us consider  a bipartite state \(\rho_{AB}\) shared by Alice and Bob. 
Perfect cloning by local operations is not possible in general. See Eqs. (\ref{debu}) and (\ref{equal_copy}))
for the meaning of this statement. 
In particular, we do \emph{not} necessarily want to 
 produce \(\varrho_{AB} \otimes \varrho_{A^{'}B^{'}}\) in a local cloning.
Since  perfect local cloning is not possible,  one may like to obtain an approximate version of it. 
And the relative entropy distance of the state that we did like to obtain in 
the (generally impossible) perfect local cloning
from that in the best approximate version, 
is defined as the ``entanglement of cloning'' 
of the state \(\varrho_{AB}\). 

Let us make everything more precise. First of all, the relative entropy distance of \(\varrho\) from 
the state \(\sigma\), denoted as \(S(\varrho|\sigma)\), is defined as 
\[
S(\varrho|\sigma) = \tr(\varrho \log_2 \varrho - \varrho \log_2 \sigma).  
\]

To begin, Alice and Bob produces a ``blank'' state \(\varrho^b_{A^{'}B^{'}}\), by LOCC. This
blank state must be a separable state, but can depend on the state \(\varrho_{AB}\). Let 
\(L\) be an LOCC map, that is applied on the input \(\varrho_{AB} \otimes \varrho^b_{A^{'}B^{'}}\), in the 
\(AA^{'}:BB^{'}\) cut. Perfect cloning requires that we have 
\begin{equation}
\label{apod}
\tr_{AB} L(\varrho_{AB} \otimes \varrho^b_{A^{'}B^{'}}) = \varrho_{A^{'}B^{'}}
\end{equation}
and 
\begin{equation}
\label{bidai}
\tr_{A^{'}B^{'}} L(\varrho_{AB} \otimes \varrho^b_{A^{'}B^{'}}) = \varrho_{AB}.
\end{equation}

This is not possible in general.  However one may try to perform it approximately in a way such that 
\begin{eqnarray}
\label{hou}
S(\varrho_{AB} | \tr_{A^{'}B^{'}} L(\varrho_{AB} \otimes \varrho^b_{A^{'}B^{'}})) \nonumber \\
+ 
S(\varrho_{A^{'}B^{'}} | \tr_{AB} L(\varrho_{AB} \otimes \varrho^b_{A^{'}B^{'}}))
\end{eqnarray}
is as small as possible. Moreover we require symmetry, i.e. the produced copies are equal:
\begin{equation}
\label{chhoi}
\tr_{A^{'}B^{'}} L(\varrho_{AB} \otimes \varrho^b_{A^{'}B^{'}})
=
\tr_{AB} L(\varrho_{AB} \otimes \varrho^b_{A^{'}B^{'}}).
\end{equation}
In that case, the quantity that we want to be as small as possible is 
\[
S(\varrho_{AB} | \tr_{A^{'}B^{'}} L(\varrho_{AB} \otimes \varrho^b_{A^{'}B^{'}})), \nonumber \\
\]
with the condition that Eq. (\ref{chhoi}) must be satisfied.
 And we define (with Eq. (\ref{chhoi}) still to be satisfied) 
\begin{equation}
\label{def_clo}
{\cal C}_E (\varrho_{AB})= \inf_L 
S( \varrho_{AB} | \tr_{A^{'}B^{'}} L(\varrho_{AB} \otimes \varrho^b_{A^{'}B^{'}}))
\end{equation}
as the ``entanglement of cloning'' of the state \(\varrho_{AB}\).
The infimum, denoted as ``inf'' in the above equation, 
is taken over all LOCC operations. In the following, we will denote 
\(\tr_{A^{'}B^{'}} L(\varrho_{AB} \otimes \varrho^b_{A^{'}B^{'}})\) by 
\({\cal L}(\varrho_{AB})\). We will call \({\cal L}\) as a ``cloning map''.


Although the measure \({\cal C}_E\)  involves a minimization 
over a relative entropy distance, it is 
very different from the 
relative entropy of entanglement \cite{VPRK, VP} of the state \(\rho_{AB}\), 
 as defined in Eq. (\ref{relent_def}).  
There is an obvious 
difference in the motivation.
Moreover 
in the case of relative entropy of entanglement, the infimum is taken
over a set of \emph{states} (for example, separable states in the case in Eq. (\ref{relent_def})), whereas
in the case of \({\cal C}_E\), the infimum is taken over a set of  \emph{operations}.

\subsubsection{Some properties} 

Let us now discuss 
some properties of \({\cal C}_E\).


(a) \emph{\({\cal C}_E\) vanishes for separable states.}
For a separable state \(\varrho_{AB}\), 
\({\cal C}_E (\rho_{AB}) =0 \)
by the very definition of \({\cal C}_E\), 
as separable states can be produced locally. 

(b) \emph{Let us now deal with the monotonicity of the measure \({\cal C}_E\)
under LOCC operations.} 
We will 
prove that 
\begin{itemize}
\item[(i)] \({\cal C}_E\) is invariant under local unitaries,
\item[(ii)] \({\cal C}_E\) is nonincreasing under addition of local ancillas.
\end{itemize}
We conjecture that 
\({\cal C}_E\) is also nonincreasing under throwing out part of the system.

Let us first show that \({\cal C}_E\) is invariant under local unitaries, 
that is
\begin{equation}
\label{saat}
{\cal C}_E (\varrho_{AB}) = {\cal C}_E(U_A \otimes U^{'}_B \varrho_{AB} U^{\dagger}_A \otimes U^{'\dagger}_B),
\end{equation}
for an arbitrary unitary \(U_A\) (\(U^{'}_B\)) on \({\cal H}_A\) (\({\cal H}_B\)). 
We have (where we leave out the suffix \(AB\) from \(\varrho_{AB}\), and denote the \emph{local}
unitary  \(U_A \otimes U^{'}_B\) by \(U\))
\begin{eqnarray}
\label{proman_unitary}
 {\cal C}_E(\varrho) & = & \inf_L S(\varrho|{\cal L}(\varrho)) \nonumber \\
          & = & \inf_L S(U \varrho U^{\dagger}|U {\cal L}(\varrho)U^{\dagger}) \nonumber \\
          & = & \inf_L S(U \varrho U^{\dagger}|U {\cal L}(U^{\dagger}U \varrho U^{\dagger}U)U^{\dagger}) \nonumber \\
           & \leq & {\cal C}_E(U \varrho U^{\dagger}). 
\end{eqnarray}
The first equality is the definition of \({\cal C}_E\). The second equality is a property of 
relative entropy distance, while the third is due to the fact that \(U\) is unitary. 
To obtain final inequality, we use the fact that \({\cal L}^{'}(\cdot) = U {\cal L}(U^\dagger \cdot U) U^\dagger\)  
is again a cloning map, for a local \(U\). 
To get the opposite inequality, \({\cal C}_E(\varrho) \geq  {\cal C}_E(U \varrho U^{\dagger})\), 
just replace \(U\) by \(U^\dagger\), and then \(\varrho\) by 
\(U \varrho U^{\dagger}\) in eq. (\ref{proman_unitary}). Thus we have obtained the relation (\ref{saat}).


Let us next 
show that \({\cal C}_E\) is nonincreasing under addition of local ancillas. 
Suppose  that a separable ancilla \(\varrho^{sep}_{A_1B_1}\) is added to the original shared state \(\varrho_{AB}\).
\(AA_1\) is with Alice, while \(BB_1\) is with Bob. Then we have (where we leave out the suffixes 
\(AB\) and \(A_1B_1\))
\begin{eqnarray}
\label{aat}
{\cal C}_E (\varrho \otimes \varrho^{sep}) & = &
\inf_L S(\varrho \otimes \varrho^{sep} | {\cal L}(\varrho \otimes \varrho^{sep}))
                                                                                                   \nonumber \\
& \leq & 
S(\varrho \otimes \varrho^{sep} | {\cal L}^{'} \otimes I (\varrho \otimes \varrho^{sep}))
                                                                                           \nonumber \\
& = & S(\varrho | {\cal L}^{'}(\varrho))                         \nonumber \\
& = & \inf_L S(\varrho | {\cal L}(\varrho))        \nonumber \\
& = & {\cal C}_E (\varrho),
\end{eqnarray}
where \({\cal L}^{'}\) is the local cloning map at which the infimum of \({\cal C}_E (\varrho_{AB})\) is 
attained, and \(I\) in the second line is the identity operator on the \(A_1B_1\) part of the system. 
The first and last equalities in Eq. (\ref{aat}) are by the definition of 
\({\cal C}_E\), while the inequality in the second line follows from 
the definition of infimum. The third line follows from the definition of relative entropy distance.
The fourth line is a consequence of the choice before, that \({\cal L}^{'}\) is the cloning map 
at which the infimum of \({\cal C}_E (\varrho_{AB})\) is 
attained.

(c) \emph{The measure \({\cal C}_E\) is a lower bound of relative entropy of entanglement.}
Suppose that for relative entropy of entanglement  of the state \(\varrho_{AB}\) (as defined 
in Eq. (\ref{relent_def}), the minimum
 is attained for the separable
state \(\sigma_{AB}\). So we have
\begin{eqnarray}
\label{tinsho}
{\cal C}_{E}(\varrho_{AB}) 
& = & \inf_{L} S( \varrho_{AB} | {\cal L}(\varrho_{AB}) ) \nonumber \\
              & \leq & S( \varrho_{AB} | \sigma_{AB} ) \nonumber \\
                   & = &   E_R(\varrho_{AB}).
\end{eqnarray}
The inequality holds as separable states can be prepared locally.  






\subsection{The entanglement of deleting}
\label{subsec_entdel}

Just as the notion of no cloning gave rise to the entanglement of cloning, 
motivated in a similar way, we will now define a measure, the 
entanglement of deleting, based on the nonexistence of a
 deleting machine.

Note that both the measures are operationally motivated.

The notion of deleting being in a sense dual to that of cloning, 
we call these two measures as dual to each other.

\subsubsection{The  definition}

As we have stressed in Sections \ref{section_forgetting} and \ref{sec_deleting_global},
in case of local deleting, the relevant operations are closed local operations, which are just 
local unitaries.

Perfect deleting is not possible in general 
by closed local operations. (See Eqs. (\ref{arai}), (\ref{poune-tin}), (\ref{tin}).)
One may however try to perform it approximately. 
And just as in the definition of entanglement of cloning, entanglement of deleting of a bipartite state
\(\varrho_{AB}\) is defined as the relative entropy distance of the state given by the (generally impossible) 
perfect local deleting machine from that in the best approximate version.

To make it precise, suppose that Alice and Bob are given two copies of the bipartite state \(\varrho_{AB}\), 
so that the input is \(\varrho_{AB} \otimes \varrho_{A^{'}B^{'}}\), with 
\(AA^{'}\) being with Alice and \(BB^{'}\) being with Bob. They apply closed local operations in the 
\(AA^{'}:BB^{'}\) cut. Suppose that the applied operations are the unitary operations \(U_{AA^{'}}\)
and \(U_{BB^{'}}\) at Alice and Bob respectively. Ideally they did like to have 
\[
\tr_{AB}U_{AA^{'}} \otimes U_{BB^{'}} \varrho_{AB} \otimes \varrho_{A^{'}B^{'}} 
U^\dagger_{AA^{'}} \otimes U^\dagger_{BB^{'}}
\]
as a separable state, while 
\[
\tr_{A^{'}B^{'}}U_{AA^{'}} \otimes U_{BB^{'}} \varrho_{AB} \otimes \varrho_{A^{'}B^{'}} 
U^\dagger_{AA^{'}} \otimes U^\dagger_{BB^{'}} = \varrho_{AB}.
\]

This being not possible in general, they may try to obtain a machine that produces a state 
in the \(A^{'}B^{'}\) part that is as ``close'' to separable as possible, and a state in the \(AB\) part that is 
as ``close'' to \(\varrho_{AB}\) as possible.
The best approximate local deleting machine is defined to be one that minimizes the quantity (where 
we have denoted the \emph{local} unitary \(U_{AA^{'}} \otimes U_{BB^{'}}\) by \(U\)) 
\begin{eqnarray}
S(\varrho_{AB} | \tr_{A^{'}B^{'}} U \varrho_{AB} \otimes \varrho_{A^{'}B^{'}} 
U^\dagger) \nonumber \\
+ \min_{\sigma_{A^{'}B^{'}} \in {\cal S}}
S(\sigma_{A^{'}B^{'}} | \tr_{AB} U \varrho_{AB} \otimes \varrho_{A^{'}B^{'}} 
U^\dagger),\nonumber 
\end{eqnarray}
where \({\cal S}\) is the set of all separable states on \({\cal H}_{A^{'}} \otimes {\cal H}_{B^{'}}\).

And the ``entanglement of deleting'' of the bipartite state \(\varrho_{AB}\) is defined as (where
\(U\) is of the form \(U_{AA^{'}} \otimes U_{BB^{'}}\))
\begin{eqnarray}
{\cal D}_E (\varrho_{AB}) = \inf_{U} \frac{1}{2}\Big[
S(\varrho_{AB} | \tr_{A^{'}B^{'}} U \varrho_{AB} \otimes \varrho_{A^{'}B^{'}} 
U^\dagger) \nonumber \\
+ \min_{\sigma_{A^{'}B^{'}} \in {\cal S}}
S(\sigma_{A^{'}B^{'}} | \tr_{AB} U \varrho_{AB} \otimes \varrho_{A^{'}B^{'}} 
U^\dagger)
\Big].
\end{eqnarray}

Note that just as in the case of entanglement of cloning, the entanglement of deleting also
involves an optimization over a set of operations.

\subsubsection{A change in the definition}
\label{subsubsec_subtle}

There is a subtlety involved in the definition of the entanglement of deleting.
The main idea of the definition of entanglement of deleting (as also for entanglement of cloning) 
is that deleting is possible globally, while not possible locally. The distance from 
what we ideally want locally from what is actually possible locally is the entanglement of 
deleting. However we must be sure that what we want ideally is possible globally or else we
will have ``some amount'' of distance from the global impossibility also, and which 
part has nothing to do with quantum correlations. 

In the definition of \({\cal D}_E\), we take a minimization over all separable states in 
\({\cal H}_{A^{'}} \otimes {\cal H}_{B^{'}}\). However this is not fair. This is because, under 
\emph{closed} global  operations, only those separable states are reachable which have the same spectrum 
as \(\varrho_{A^{'}B^{'}}\). 

Henceforth, whenever we consider \({\cal D}_E\) for a state \(\varrho_{AB}\), we will keep 
in mind that the minimization in the definition is over such separable states. However, for 
brevity, we will continue to say that the minimization is over ``separable states on  
\({\cal H}_{A^{'}} \otimes {\cal H}_{B^{'}}\)''. 

So for example, if \(\varrho_{AB}\) is pure, the minimization is over all pure product states. This 
consideration will be important when we obtain a bound on \({\cal D}_E\) for pure bipartite states, 
in Section \ref{subsubsection_delete_bound}.

\subsubsection{Some properties}

Let us now discuss some properties of the entanglement of deleting. 

(a) \emph{The entanglement of deleting vanishes for separable states.} This follows directly from the 
definition of \({\cal D}_E\).

(b) \emph{We now consider the monotonicity of \({\cal D}_E\) under LOCC operations.} 
We will show that 
\begin{itemize}
\item[(i)] \({\cal D}_E\) is invariant under local unitary operations,
\item[(ii)] \({\cal D}_E\) is nonincreasing under addition of local ancillas.
\end{itemize}
It seems probable that \({\cal D}_E\) is also nonincreasing under tracing out part of the system.

Let us first prove item (i). For any \emph{local} unitary operator \(V\) on 
 \({\cal H}_A \otimes {\cal H}_B\),  we have  
\begin{widetext}
\begin{eqnarray}
{\cal D}_E (\varrho_{AB}) = \inf_{U} \frac{1}{2}\Big[
             S(\varrho_{AB} | \tr_{A^{'}B^{'}} U \varrho_{AB} \otimes \varrho_{A^{'}B^{'}} U^\dagger) 
      +      \min_{\sigma_{A^{'}B^{'}} \in {\cal S}}
           S(\sigma_{A^{'}B^{'}} | \tr_{AB} U \varrho_{AB} \otimes \varrho_{A^{'}B^{'}} U^\dagger) 
													\Big] \nonumber \\
= \inf_{U} \frac{1}{2}\Big[
    S(V \varrho_{AB} V^\dagger | V \tr_{A^{'}B^{'}} \{ U \varrho_{AB} \otimes \varrho_{A^{'}B^{'}} U^\dagger \} V^\dagger) 
      +      \min_{\sigma_{A^{'}B^{'}} \in {\cal S}}
      S( V \sigma_{A^{'}B^{'}} V^\dagger | V \tr_{AB}\{U \varrho_{AB} \otimes \varrho_{A^{'}B^{'}} U^\dagger \} V^\dagger) 
													\Big] \nonumber \\
= \inf_{U} \frac{1}{2}\Big[
    S \left( V \varrho_{AB} V^\dagger | 
 \tr_{A^{'}B^{'}} \left\{ (V \otimes I_{A^{'}B^{'}}) U (V^\dagger \otimes V^\dagger)
(V \varrho_{AB} V^\dagger)
 \otimes (V \varrho_{A^{'}B^{'}} V^\dagger)  (V \otimes V) U^\dagger (V^\dagger \otimes I_{A^{'}B^{'}}) \right\} \right) 
 																																					\nonumber \\
      +      \min_{\sigma_{A^{'}B^{'}} \in {\cal S}}
      S \left( V \sigma_{A^{'}B^{'}} V^\dagger | 
 \tr_{AB} \left\{ (I_{AB} \otimes V) U  (V^\dagger \otimes V^\dagger) 
(V \varrho_{AB} V^\dagger) \otimes (V \varrho_{A^{'}B^{'}}V^\dagger) (V \otimes V)
 U^\dagger 
(I_{AB} \otimes V^\dagger) \right\}  \right) 
													\Big] \nonumber \\
\leq {\cal D}_E (V \varrho_{AB} V^\dagger). 
\end{eqnarray}
\end{widetext}
Note that the same \(V\), which is defined on \({\cal H}_A \otimes {\cal H}_B\), is also defined on 
\({\cal H}_{A^{'}} \otimes {\cal H}_{B^{'}}\), as \(\dim {\cal H}_A = \dim {\cal H}_{A^{'}}\) and 
\(\dim {\cal H}_B = \dim {\cal H}_{B^{'}}\), since we have two copies of the same state as input, in a 
deleting procedure. We have here used the facts that for any unitary operator \(U\),
\(S(\varrho|\sigma) = S(U \varrho U^\dagger | U \sigma U^\dagger)\), and that \(V\) is unitary.
Also we have used the relation that 
\(V_{AB} (\tr_{CD} \eta_{ABCD}) V_{AB}^\dagger = 
\tr_{CD} (V_{AB} \otimes I_{CD}) \eta_{ABCD} (V_{AB}^\dagger \otimes I_{CD})\). This relation actually holds for any 
operation on the \(AB\) part.

Let us next prove item (ii). Suppose that a local ancilla \(\varrho_{A_1B_1}^{sep}\) is added to the state
\(\varrho_{AB}\), with the \(AA_1\) part being with Alice and the \(BB_1\) part with Bob. Since 
\(\varrho_{A_1B_1}^{sep}\) is produced locally, it is a separable state. Let 
\(U^{'}_{AA^{'}} \otimes U^{''}_{BB^{'}}\) be the local unitary operation, and 
let \(\sigma^{'}_{A^{'}B^{'}}\) be the separable state at which 
the infimum of \({\cal D}_E(\varrho_{AB})\) is attained. We have (where 
\(U\) is now a \emph{local} unitary operator in the \(AA_1A^{'}A^{'}_1 : BB_1B^{'}B^{'}_1\) cut, and 
\({\cal S}\) is now the set of separable states in the \(A^{'}A^{'}_1 : B^{'}B^{'}_1\) cut) 
\begin{widetext}
\begin{eqnarray}
{\cal D}_E(\varrho_{AB} \otimes \varrho^{sep}_{A_1B_1})
= \inf_U \frac{1}{2} \Big[
S(\varrho_{AB} \otimes \varrho^{sep}_{A_1B_1} | \tr_{A^{'}B^{'}A_1^{'} B_1^{'}}
 U \varrho_{AB} \otimes \varrho^{sep}_{A_1B_1} \otimes \varrho_{A^{'}B^{'}} \otimes \varrho^{sep}_{A_1^{'}B_1^{'}} 
U^\dagger)  \nonumber \\
+ \min_{\sigma_{A^{'}B^{'}A_1^{'} B_1^{'}} \in {\cal S}}
S(\sigma_{A^{'}B^{'}A_1^{'} B_1^{'}} | \tr_{ABA_1B_1} U \varrho_{AB} \otimes \varrho^{sep}_{A_1B_1} 
\otimes \varrho_{A^{'}B^{'}} \otimes \varrho^{sep}_{A^{'}_1B^{'}_1} 
U^\dagger)
\Big] \nonumber \\
\leq \frac{1}{2} \Big[ 
S \left( \varrho_{AB} \otimes \varrho^{sep}_{A_1B_1} | \tr_{A^{'}B^{'}A_1^{'} B_1^{'}} \left\{
 (U^{'}_{AA^{'}} \otimes U^{''}_{BB^{'}}
 \varrho_{AB} \otimes \varrho_{A^{'}B^{'}}  U^{'\dagger}_{AA^{'}} \otimes U^{''\dagger}_{BB^{'}})
 \otimes \varrho^{sep}_{A_1B_1}  \otimes \varrho^{sep}_{A_1^{'}B_1^{'}} 
 \right\} \right)  \nonumber \\
+ 
S \left( \sigma^{'}_{A^{'}B^{'}} \otimes \varrho^{sep}_{A_1^{'} B_1^{'}} | 
\tr_{ABA_1B_1} 
\left\{
 (U^{'}_{AA^{'}} \otimes U^{''}_{BB^{'}}
 \varrho_{AB} \otimes \varrho_{A^{'}B^{'}}  U^{'\dagger}_{AA^{'}} \otimes U^{''\dagger}_{BB^{'}})
 \otimes \varrho^{sep}_{A_1B_1}  \otimes \varrho^{sep}_{A_1^{'}B_1^{'}} 
 \right\}
\right)\Big]  = {\cal D}_E(\varrho_{AB}). \nonumber 
\end{eqnarray}
\end{widetext}
The inequality follows by choosing \(U\) as \(U^{'}_{AA^{'}} \otimes U^{''}_{BB^{'}} \otimes I_{A_1A_1^{'}B_1B_1^{'}}\),
and \(\sigma_{A^{'}B^{'}A_1^{'} B_1^{'}}\) as \(\sigma^{'}_{A^{'}B^{'}} \otimes \varrho^{sep}_{A_1^{'} B_1^{'}}\)
in the infimum of \({\cal D}_E(\varrho_{AB} \otimes \varrho^{sep}_{A_1B_1})\). The last 
equality is by straightforward calculation.

\section{Bounds for pure states}
\label{section_bounds}

In this section, we will obtain bounds of our measures of entanglement for the case of pure states.

\subsection{Bound on entanglement of cloning for pure states}


For definiteness, let us consider the case of two qubits.


To obtain the bound, we take advantage of the fact that a cloning machine is known, that 
produces two symmetric inexact copies out of a \emph{single unknown} qubit optimally
\cite{Bruss-cloning} (see also \cite{Buzek}).  
The unitary transformation of the cloning machine is defined on a subspace of the space of three qubits,
one of which (the third qubit in Eq. (\ref{cloning_map}) below)
acts as a sort of an environment, while the other two are the two symmetric copies. The 
transformation (\(U^{'}\)) is defined as follows:
\begin{eqnarray}
\label{cloning_map}
 U^{\prime }\left| 0\right\rangle
\left| b\right\rangle \left| e_1\right\rangle
=\sqrt{\frac{2}{3}}\left| 00\right\rangle \left| e\right\rangle
+\sqrt{\frac{1}{6}}(\left| 01\right\rangle +\left| 10\right\rangle
)\left| e_{\perp }\right\rangle \nonumber \\
U^{\prime }\left| 1\right\rangle \left| b\right\rangle
\left| e_1\right\rangle =\sqrt{\frac{2}{3}}\left| 11\right\rangle
\left| e_{\perp }\right\rangle +\sqrt{\frac{1}{6}}(\left|
01\right\rangle +\left| 10\right\rangle )\left| e\right\rangle,
\end{eqnarray}
where \( \left| b\right\rangle  \) is a \emph{fixed} blank state
(in a two-dimensional Hilbert space), \( \left| e_1\right\rangle  \)
is the initial state of the environment, 
\( \left| e\right\rangle  \)
and \( \left| e_{\perp }\right\rangle  \) being two mutually
orthonormal states of the environment Hilbert space. The two clones
are to surface at the first and second qubits. Note that the
environment has turned out to be a qubit.

In our situation, Alice and Bob are given a \(2 \otimes 2\) state \(\varrho_{AB}\).
They want to produce two copies of it, in the sense discussed earlier (see Eqs. (\ref{apod}), (\ref{bidai}),
(\ref{hou})). To use the unitary operator \(U^{'}\), they prepare local blank states and local environment 
states and then both of them separately apply \(U^{'}\), to obtain
\begin{equation}
\label{dusho}
U^{'} \otimes U^{'} \varrho_{AB} 
(\left|b\right\rangle \left\langle b \right|)_{A^{'}} 
(\left|b\right\rangle \left\langle b \right|)_{B^{'}} 
(\left|e\right\rangle \left\langle e \right|)_{A_e} 
(\left|e\right\rangle\left\langle e \right|)_{B_e}
U^{'\dagger} \otimes U^{'\dagger},
\end{equation}
where the \(AA^{'}A_e\) part is with Alice and the \(BB^{'}B_e\) part is with Bob, so that the first 
\(U^{'}\) is applied on \(AA^{'}A_e\), and the second on \(BB^{'}B_e\). 
Then Alice performs a swap operation on the \(AA^{'}\) part of her Hilbert space. (This 
swap could obviously have been performed by either Alice or Bob.) Alice and Bob also 
throws out the environmental parts now (this could of course have been done before the swap).
This will give us a local cloning machine and its performance will be a bound on the 
performance of the best local cloning machine.

Let us calculate the bound when Alice and Bob shares 
a general pure state \(\left|\psi_{AB}\right\rangle = 
a \left|0 0\right\rangle_{AB} + b \left|1 1 \right\rangle_{AB}\), shared by Alice and Bob,  
in \(2 \otimes 2\). Here \(a\) and \(b\) are positive, and \(a^2 + b^2 =1\). 

In that case, after the local cloning operation (adding 
the local blank and environment states, applying \(U^{'} \otimes U^{'}\), swapping \(A\) with \(A^{'}\), and 
then tracing out the environments) has been performed, 
the state obtained in the \(AB\) part (which is 
equal to that in the \(A^{'}B^{'}\) part) is \cite{ekso,sto}
\begin{eqnarray}
 \varrho^{clone}_{AB} = \varrho_{A^{'}B^{'}}^{clone} = 
\frac{24a^2+1}{36}\left|00\right\rangle \left\langle 00 \right|
+ \frac{24b^2+1}{36}\left|11\right\rangle \left\langle 11 \right| \nonumber \\
+ \frac{5}{36} (\left|01\right\rangle \left\langle 01 \right|
+  \left|10\right\rangle \left\langle 10 \right|)
+ \frac{4ab}{9} (\left|00\right\rangle \left\langle 11 \right|
+ \left|11\right\rangle \left\langle 00 \right|). \nonumber \\
\end{eqnarray}

Our bound on \({\cal C}_E\), for the case of the state 
\(\left|\psi_{AB}\right\rangle = 
a \left|0 0\right\rangle_{AB} + b \left|1 1 \right\rangle_{AB}\), is therefore given by 
\begin{eqnarray}
\label{clone-er_bound}
{\cal C}_E (\left|\psi_{AB}\right\rangle) 
\leq S(\left|\psi\right\rangle \left\langle \psi \right| | \varrho^{clone}).
\end{eqnarray}

\begin{figure}[tbp]
\begin{center}
\epsfig{figure= 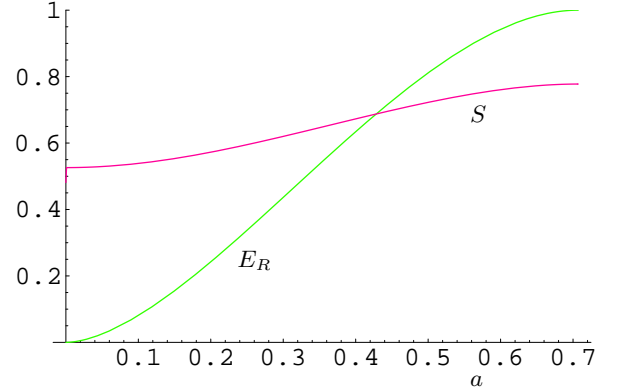,width=0.45\textwidth}
\put(-52,95){\(S\)}
\put(-140,40){\(E_R\)}
\put(-52,-5){\(a\)}
\caption{Plot of the bound \(S(\left|\psi\right\rangle \left\langle \psi \right| | \varrho^{clone})\)
in Eq. (\ref{clone-er_bound}), denoted as \(S\) in the figure, and the relative entropy 
of entanglement (\(E_R\)), for pure states 
\(\left|\psi \right\rangle = a \left|0 0\right\rangle + b \left|1 1 \right\rangle\).
Both are upper bounds of \({\cal C}_E\), so that the  curve for \({\cal C}_E\) 
must lie below both the curves in the figure. 
The right end of the horizontal axis corresponds to 
the maximally entangled state (in \(2 \otimes 2\)) 
\(\frac{1}{\sqrt{2}}(\left|0 0\right\rangle + \left|1 1 \right\rangle)\). Note that 
the bound rules out the possibility of \({\cal C}_E =1\) for this state.  \({\cal C}_E =1\) 
arises naturally  for the maximally entangled state in \(2 \otimes 2\), for almost all entanglement measures.
} 
\end{center}
\label{satyajit}
\end{figure}

In Fig. 2,
we plot this 
bound of \({\cal C}_E\) for the case of pure states.
For pure states, the von Neumann entropy of either local density martix is known to be the ``good 
asymptotic'' measure of entanglement \cite{IBMpure,DHR}. Let us call it \(E(\left|\psi\right\rangle)\).
Therefore \(E(\left|\psi_{AB}\right\rangle)= 
S(\tr_A(\left|\psi\right\rangle \left\langle \psi \right|)_{AB})\), 
where \(S(\varrho)= -\tr \varrho \log_2 \varrho\) is the 
von Neumann entropy of \(\varrho\). 
We compare our bound for \({\cal C}_E\)
in Eq. (\ref{clone-er_bound}),
with \(E(\left|\psi\right\rangle)\). Moreover note that 
\({\cal C}_E\) is for all states bounded above by relative entropy of entanglement \(E_R\) 
(see Eq. (\ref{tinsho})), and 
\(E_R\) for pure states coincides with \(E(\left|\psi\right\rangle)\) \cite{VP}. 
Thus the bound that we have obtained for 
pure states is the minimum of \(E(\left|\psi\right\rangle) = E_R(\left|\psi \right\rangle)\) and 
the bound in Eq. (\ref{clone-er_bound}):
\begin{equation}
\label{fulkopi}
{\cal C}_E(\left|\psi\right\rangle) \leq 
\min\{E_R(\left|\psi\right\rangle), S(\left|\psi\right\rangle \left\langle \psi \right| | \varrho^{clone})\}
\equiv {\cal C}_E^{bound}.
\end{equation}
The  figure (Fig. 2)
shows that the bound in  Eq. (\ref{clone-er_bound})
is better than the relative entropy of entanglement bound, for 
high entangled states.  For low entangled states,
the relative entropy of entanglement bound is better.
The change-over of the bound from relative entropy of entanglement to 
the bound in Eq. (\ref{clone-er_bound})
occurs at \(a\approx 0.4282\).

For the maximally entangled state in \(2 \otimes 2\), we have 
\begin{equation}
\label{bloom}
{\cal C}_E \leq \log_2\frac{12}{7} \approx 0.777608.
\end{equation}
The cloning machine of Eq. (\ref{cloning_map}) has the property of being ``isotropic''.
That is, the quality of the clones at the output is equally good, irrespective the input state.  
It therefore seems that the local cloning machine considered here to obtain the bound in Eq. (\ref{clone-er_bound})
is optimal (i.e. it 
attains the infimum in \({\cal C}_E\)), for the maximally entangled states in \(2 \otimes 2\).

Similar upper bounds can be 
obtained for higher dimensional 
pure states, by considering higher dimensional cloning machines.

\subsection{Bound on entanglement of deleting for pure states}
\label{subsubsection_delete_bound}

We now try to obtain a bound on the entanglement of deleting for the case of pure states, where again 
for definiteness, we consider only states in \(2 \otimes 2\).

The problem here is that an optimal deleting machine for a single system is not known. Even if 
we find one, it will be defined only on the parallel space (see \cite{not_really}). However
if Alice and Bob are given two copies of \(\left|\psi \right\rangle = a \left| 00 \right\rangle
+ b\left| 11 \right\rangle\), the shared state is 
\begin{equation}
\label{ebar}
(a \left| 00 \right\rangle
+ b\left| 11 \right\rangle)_{AB} 
\otimes 
(a \left| 00 \right\rangle
+ b \left| 11 \right\rangle)_{A^{'}B^{'}},
\end{equation}
with \(AA^{'}\) being with Alice and \(BB^{'}\) with Bob. 
And then the \(AA^{'}\) part (as also the 
\(BB^{'}\) part) is the whole four-dimensional Hilbert space, instead of being just the 
parallel space.

As a way out, we consider the local deleting machine that swaps \(A\) with \(A^{'}\) at Alice's end. 
(The swapping could obviously have been done at Bob's end also.)  In that case, if 
the state in Eq. (\ref{ebar}) is the input to the deleting machine, we have 
(\(a\) and \(b\) are positive, and \(a^2 + b^2 =1\))
\[
\varrho^{delete} = a^2 \left| 0 \right\rangle \left\langle 0 \right|
+ b^2 \left| 1 \right\rangle \left\langle 1 \right|
\otimes a^2 \left| 0 \right\rangle \left\langle 0 \right|
+ b^2 \left| 1 \right\rangle \left\langle 1 \right|
\]
as the output at both \(AB\) and \(A^{'}B^{'}\). 

The quality of this local deleting machine for pure entangled inputs, is 
\[
S(\left\langle \psi \right| \left\langle \psi \right| | \varrho^{delete}) 
+ \min S(\varrho^{sep} | \varrho^{delete}),
\]
where the minimization is over all \emph{pure product}  states in \(2 \otimes 2\) 
(see Section \ref{subsubsec_subtle}). 
A half of this quantity is therefore an upper bound of the entanglement of deleting for 
pure states in \(2 \otimes 2\). Performing the minimization, we have
for pure states in \(2 \otimes 2\), 
\begin{equation}
\label{delete-er_bound}
{\cal D}_E (\left|\psi \right\rangle) \leq 
E( \left|\psi \right\rangle) - 2 \log_2 b \equiv {\cal D}_E^{bound},
\end{equation}
where we have assumed that \(b \geq a\). The minimization is attained at \(\left|11\right\rangle\).
For the maximally entangled state in \(2 \otimes 2\), we have \({\cal D}_E \leq 2\).

We plot this bound, \({\cal D}_E^{bound}\),
for entanglement of deleting in Fig. 3, where, for comparison,
 we have also plotted the bound, \({\cal C}_E^{bound}\) (Eq.
(\ref{clone-er_bound})),  for 
entanglement of cloning obtained in the preceeding subsection.

\begin{figure}[tbp]
\begin{center}
\epsfig{figure= 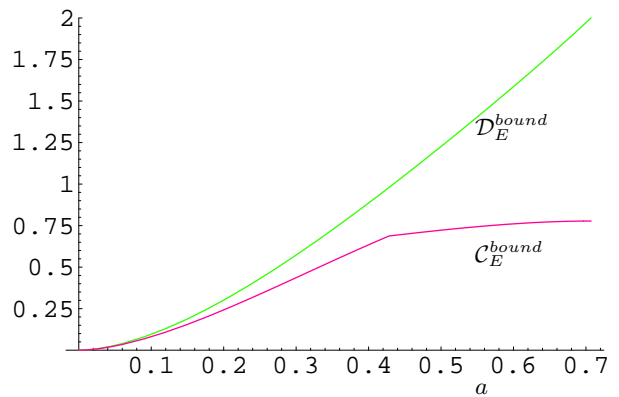,width=0.45\textwidth}
\put(-52,93){\({\cal D}_E^{bound}\)}
\put(-52,45){\({\cal C}_E^{bound}\)}
\put(-52,-5){\(a\)}
\caption{Plot of the bound \({\cal D}_E^{bound}\) in Eq. (\ref{delete-er_bound})
of entanglement of deleting,
for the pure states \(\left| \psi \right\rangle = 
a \left|0 0\right\rangle + b \left|1 1 \right\rangle\). We have assumed that \(b \geq a\). 
For comparison, we have also drawn the bound (\({\cal C}_E^{bound}\)) in Eq.
(\ref{clone-er_bound}) for the same pure states.
Therefore entanglement of deleting must lie below the upper curve, while 
entanglement of cloning must lie below the lower curve. 
So apparantly it seems that \({\cal D}_E\) is larger than \({\cal C}_E\). This is 
to be expected, as the   set of operations for the case of local 
deleting is drastically smaller than that for local cloning.
} 
\end{center}
\label{poetry}
\end{figure}

\section{Discussions}
\label{sec_discu}

Entanglement is a multi-faceted entity, and it seems that it cannot be quantized or 
understood by a single 
entanglement measure. Therefore it is interesting to quantify it in as many ways as possible. However
the measures of entanglement that has an operational meaning, like the entanglement cost or 
distillable entanglement are important in understanding the resource perspective of entanglement. 

In this paper, we have defined two entanglement measures for bipartite states which are motivated 
by the no cloning and no deleting results in quantum mechanics.

Interestingly, the entanglement of cloning (\({\cal C}_E\)) has been found to be strictly less than unity for a 
maximally entangled state in \(2 \otimes 2\) (Eq. (\ref{bloom})). 
Almost all entanglement measures turn out to be unity for the maximally entangled states in \(2 \otimes 2\).
Even the single-copy pure state entanglement transformation monotones \cite{Vidal-mon2000,Nielsen} 
of the maximally entangled states 
in \(2 \otimes 2\) are \(\frac{1}{2}\) and \(1\). Leaving out the trivial measure \(1\) (which 
is a measure even for nonmaximally entangled pure states, the other measure is just unity 
after applying logarithm (to the base 2) and a minus sign. It is interesting to find out 
whether a normalised \({\cal C}_E\), i.e. \({\cal C}_E\) divided by the value of \({\cal C}_E\) 
for a maximally entangled state in \(2 \otimes 2\) is actually different from the known entanglement 
measures, even for pure states.

Let us mention here that we do not know whether there is an order between the two proposed entangled 
measures. However, since the set of operations in local deleting is drastically reduced 
as compared to that in local cloning, it seems that entanglement of deleting will in 
general be larger than the entanglement of cloning (see Fig. 3). 

Let us also indicate here that both the proposed measures can be generalised to the case of 
\emph{multipartite} states in a straightforward manner. Such generalisation is 
interesting, as such multipartite entanglement measures will 
not quantify entanglement in the usual way. Let us try to see this for the case of 
multipartite entanglement of cloning. The bipartite as well as the multipartite entanglement 
of cloning (as well as that of deleting) will of course have a certain continuity, following 
from the continuity of relative entropy.  Consequently for states of the form 
\(a \left|000\right\rangle + b \left|111\right\rangle\), the multipartite entanglement 
of cloning must be vanishingly small when \(a\) is vanishing. This is because the 
multipartite entanglement of cloning must vanish for the state \(\left|111\right\rangle\). Three 
separated partners can produce as many copies of \(\left|111\right\rangle\) as they want, by local 
operations. However the state 
\(\left|0\right\rangle \otimes \frac{1}{\sqrt{2}}(\left|00\right\rangle
+ \left|11\right\rangle)\)
has a finite value (whatever that is) of multipartite entanglement of cloning, since this state cannot 
be produced by three separated partners by local actions. Both the proposed measures can also be 
generalised to the case of a set of states, even for the case of a single system (see \cite{jomey_doi} in this 
regard). We will later on work on these generalisations.

\begin{acknowledgments}

MH is supported by the Polish Ministry of Scientific
Research and Information Technology under the (solicited) grant No.
PBZ-MIN-008/P03/2003 and by EC grants RESQ, Contract No. IST-2001-37559 
and  QUPRODIS, Contract No. IST-2001-38877.
AS thanks S.L. Braunstein for comments in Ref. \cite{not_really}. AS and US
acknowledge support from the Alexander von Humboldt Foundation.

\end{acknowledgments}


\begin{thebibliography}{10}
\bibitem{not_really} The transformation need be unitary only on the parallel subspace, and not on 
the whole Hilbert space. Let \(\left|\psi\right\rangle\) belong to the Hilbert space 
\({\cal H}\). Then the parallel space, \({\cal P}\), 
is defined to be the subspace, of \({\cal H} \otimes {\cal H}\),
that is spanned by all vectors of the form \(\left|\psi\right\rangle \otimes \left|\psi\right\rangle\). 
In proving the no deleting theorem, that is to prove the nonexistence of a transformation that achieves 
(\ref{global_delete}), we need only 
assume that the transformation is unitary on the subspace 
\({\cal P}\).
\bibitem{ekso} This machine was also considered in Ref. \cite{sto} for the purpose of what was called 
``broadcasting of entanglement''. However note here that the concept of broadcasting of entanglement
requires some separability requirements (see \cite{sto}) on the output state 
at \(ABA^{'}B^{'}\). These separability constraints are not required in our considerations. 
It is therefore conceivable that better local cloning machines than the one in Eq. (\ref{dusho}), can 
be obtained, at least for mixed entangled states.
\end{thebibliography}
\end{document}